\documentclass[aps,prl,preprint,groupedaddress]{revtex4}
\usepackage{graphicx}
\begin{document}

%\preprint{}

\title{Coalescence constraints of many-body systems in one dimension}


\author{Xiao-Yin Pan and Viraht Sahni }
%\thanks{}
\affiliation{Department of Physics, Brooklyn College and the
Graduate School of the City University of New York, New York, New
York 10016. }


\date{\today}

\begin{abstract}
For one-dimensional many-body systems interacting via the \textit{Coulomb force} and with \textit{arbitrary} external potential energy, we derive (\textit{i}) the \textit{node coalescence condition} for the wave function. This condition rigorously proves the following:  (\textit{ii}) that the particles satisfy \textit{only} a node coalescence condition; (\textit{iii}) that irrespective of their charge or statistics, the particles cannot coalesce; (\textit{iv}) that the particles cannot cross each other, and must be ordered; (\textit{v}) the particles are  therefore distinguishable; (\textit{vi}) as such their statistics are not significant; (\textit{vii}) conclusions similar to those of the spin-statistics theorem of quantum field theory are arrived at via non-relativistic quantum mechanics; (\textit{viii}) the noninteracting system \textit{cannot} be employed as the lowest-order in a perturbation theory of the interacting system. (\textit{ix}) Finally, the coalescence condition for particles with the short-ranged delta-function interaction and \textit{arbitrary} external potential energy,  is also derived.  These particles can coalesce and cross each other.  We further note that the ordering of particles in one-dimension occurs  only for those interaction potential energies for which the wavefunction satisfies a node coalesence condition.

\end{abstract}

%\keywords{}

\maketitle

%\section{I. Introduction}
The theoretical interest in exactly solvable one-dimensional systems goes back to the beginning of quantum mechanics \cite{1}.  Such models often explain the key ideas underlying the physics, which are then generalized and refined by application to real physical systems. Thus, for example, the Kronig-Penney model \cite{1,2} elucidates the critical ideas of the band structure of solids, and of the resulting concepts of the forbidden energy gap and effective mass of the electrons. There has, however, been a recent resurgence of this theoretical interest in one-dimensional systems \cite{3,4}.   The principal motivation for this has been the experimental realization of one-dimensional systems interacting via the Coulomb force, and by their subsequent and possible future technological applications.  For example \cite{3}, semiconductor nanostructures, carbon nanotubes, inorganic and organic chains are all physical manifestations of the one-dimensional electron gas and these materials are of value in nanotechnology.  Additionally, experimental results exist, such as the measurement of terrace width distributions of crystal surfaces that too can be explained \cite{4} on the basis of one-dimensional models. The theoretical study of the fundamental properties of such systems is therefore of continuing interest.  This paper is a \textit{rigorous} description of the coalescence constraints, and of the resulting consequences, of one-dimensional many-body systems interacting via the \textit{Coulomb force}
and with \textit{arbitrary} external potential energy.
\\

  To contrast with the significantly different physics arrived at in one dimension, we review the coalescence conditions of many-body systems in dimensions $D \geq 2$.  The motion of electrons in an external field ${\bf F}^{ext}({\bf r}) = -\nabla v({\bf r})$ is correlated due to the Pauli exclusion principle and Coulomb repulsion.  Electrons of parallel spin cannot coalesce as a consequence of the exclusion principle.  This is best understood by writing the wavefunction as an infinite sum of Slater determinants.  If the spatial coordinates of two electrons with parallel spin are the same, then each Slater determinant, and hence the wavefunction vanishes.  However, electrons of antiparallel spin can coalesce inspite of the fact that they interact via the Coulomb force that is singular at coalescence. Similarly, an electron and a positively charged nucleus can coalesce again inspite of the Coulomb interaction between them being singular at coalescence.  This ability at coalescence between electrons of antiparallel spin or between an electron and a nucleus is reflected by the wavefunction satisfying the \textit{cusp} or \textit{node coalescence condition}.  The integral and differential forms \cite{5} of the coalescence conditions for $D \geq  2$ are
  \begin{equation}
   \Psi ({\bf r}_{1},{\bf r}_{2},...{\bf r}_{N})= \Psi ({\bf r}_{2},{\bf r}_{2},{\bf r}_{3},...,{\bf r}_{N})(1+\frac{2 Z_{1} Z_{2}\mu_{12}}{D-1} r_{12} )+ {\bf r}_{12}\cdot {\bf C}({\bf r}_{2},{\bf r}_{3},...,{\bf r}_{N}),
 \end{equation} 
 and
 \begin{equation}
 (\frac{ \partial {\bar \Psi}}{\partial r_{12}}) |_{r_{12}\rightarrow 0} = \frac{2 Z_{1} Z_{2}\mu_{12}}{D-1} \;\; \Psi(r_{12}=0),
\end{equation}
where $Z_{1}$ and $Z_{2}$ are the charges of particles $1$ and $2$, $r_{12} = | {\bf r}_{1} - {\bf r}_{2} |$, ${\bf r}_{12}={\bf r}_{1} - {\bf r}_{2}$, $\mu_{12} = \frac{m_{1} m_{2}}{m_{1} + m_{2}}$ ,  $m_{1}$ and $m_{2}$ the masses of the particles, ${\bf C}({\bf r}_{2},{\bf r}_{3},...,{\bf r}_{N})$ an undetermined vector, and ${\bar \Psi}$ the wavefunction spherically averaged about the point of coalescence.  For electron-nucleus coalescence, $Z_{1} = - 1, Z_{2} = Z$ the nuclear charge, and $\mu_{12}\approx m_{e}$ the mass of the electron.  For the electron-electron coalescence, $Z_{1} = - 1, Z_{2} = - 1, \mu_{12} = m_{e}/2$.  For $D = 3$, the traditional integral and differential cusp conditions are recovered \cite{6}.  Note that at coalescence the wavefunction may either exhibit a cusp or have a node.  If the wavefunction vanishes at coalescence, then the condition is referred to as the \textit{node coalescence condition}.  Otherwise it is referred to as the \textit{cusp coalescence condition}. In the $D = 3$ case, for both the electron-nucleus and electron-electron coalescence the wavefunction  usually satisfies a cusp coalescence condition.  For example, in the Hydrogen atom \textit{ground} state, the electron density at the nucleus $\rho({\bf r})|_{{\bf r} = 0}  =  \Psi^{*} \Psi|_{{\bf r} = 0}  $  which is the probability of the electron being there, is positive-definite. For the \textit{ground} state of the Hooke's atom \cite{7}, the electron pair-correlation density $g({\bf r} {\bf r}')=\left\langle \Psi|\sum'_{ij}\delta({\bf r}-{\bf r}_{i}) \delta({\bf r}-{\bf r}_{j})|\Psi\right\rangle/\rho({\bf r})$ which is the the conditional probability density at ${\bf r}'$ for an electron at ${\bf r}$, is also positive-definite \cite{8} at ${\bf r} ={\bf r}' $. ( The Hooke's atom is comprised of two electrons interacting via the Coulomb force, but whose external potential energy due to the nucleus of charge $Z = 2$ is harmonic. For certain discrete values of the spring constant, the wavefunction is known in closed analytical form.) Thus, in the Hydrogen atom case, the electron can cross over the nucleus, and in the case of the Hooke's atom, the two electrons can cross each other. On the other hand, the wavefunction of the Hydrogen atom in a $p$ state satisfies the node coalescence condition for electron-nucleus coalescence.  As another example, in $D = 2$, the approximate Laughlin wavefunction \cite{9} for the fractional Quantum Hall Effect satisfies \cite{5} the node coalescence condition.\\

  In this paper we prove the following results and conclusions for quantal particles in $D = 1$ dimension space interacting through the \textit{Coulomb force} and with \textit{arbitrary} external potential energy $v({\bf r})$. (\textit{i} )  We have derived the \textit{node coalescence condition} for the wavefunction; (\textit{ii}) As such, the particles satisfy \textit{only} a node coalescence condition; ( \textit{iii} ) Irrespective of their charge or statistics, these particles \textit{cannot} coalesce. This \textit{local} property of non-coalescence of particles is valid irrespective of the topology of the one-dimensional system.  For example, the particles could be confined in a ring. (That identical particles with parallel spin cannot coalesce in $D = 1$ space also follows from the Pauli exclusion principle. Here we prove the more \textit{general} result that \textit{any} two particles interacting through the Coulomb force cannot coalesce in $D = 1$ dimension space.); ( \textit{iv} ) Therefore, the particles cannot cross each other and must be ordered; ( \textit{v} ) Hence the particles are distinguishable; ( \textit{vi} ) Thus, in $D = 1$ space, the statistics of the particles are not significant; (\textit{vii} ) Item ( \textit{vi} ) is akin to the $D = 1$ spin-statistics theorem of quantum field theory.  Here we have arrived at similar conclusions via non-relativistic quantum mechanics; ( \textit{viii} ) The statistics of \textit{noninteracting} particles in $D = 1$ space \textit{are} of significance: their energy spectrum will differ depending on whether the particles are bosons or fermions.  Consequently, in $D = 1$ space, the noninteracting system cannot be employed as the lowest-order approximation in a perturbation theory of the interacting system; (\textit{ix} ) To contrast with the case of particles interacting via the  Coulomb force, we have also derived the $D = 1$ coalescence condition for the short-ranged delta-function interaction.  In the latter case, as is known, the particles can coalesce and cross each other.\\

  We begin by deriving the node coalescence condition for the wavefunction.
  The nonrelativistic  Schr{\" o}dinger equation for $N$ charged particles in $D=1$ space  is 
  \begin{equation}
 {\hat H} \Psi (x_{1},x_{2},...x_{N})=E \Psi (x_{1},x_{2},...,x_{N}),
 \end{equation} 
  where the Hamitonian ${\hat H} $ 
  is
\begin{equation}
 {\hat H}= -\sum_{i=1}^{N} \frac{1}{2 m_{i}} \frac{\partial ^{2}}{\partial x_{i}^{2}} +\sum_{j>i=1}^{N} \frac{Z_{i} Z_{j}}{|x_{i}-x_{j}|}+\sum_{i=1}^{N} v(x_{i}),
\end{equation}  
 $m_{i}$ and $Z_{i}$ are the mass and charge of the \textit{i}th particle, and $ v(x_{i})$
an arbitrary  external potential energy. Focus on any two particles, say $1$ and $2$.
We are interested in the behaviour of the wave function when the distance between them becomes very small. First transform the coordinates $x_{1}$ and  $x_{2}$ to their center of mass $X_{12}$ and relative $x_{12}$ coordinates:
\begin{equation}
   X_{12}  =  \frac{m_{1} x_{1}+m_{2} x_{2}}{m_{1}+m_{2}}, %\nonumber \\
  \end{equation}
  \begin{equation}
   x_{12}  =  x_{1}-x_{2},
   \end{equation}
 so that
   \begin{equation}
   -\frac{1}{2 m_{1}} \frac{\partial^{2}}{\partial x_{1}^{2}}- \frac{1}{2 m_{2}} \frac{\partial^{2}}{\partial x_{2}^{2}} = - \frac{1}{2 (m_{1}+m_{2})}  \frac{\partial^{2}}{\partial X_{12}^{2}} - \frac{1}{2 \mu_{12}}  \frac{\partial^{2}}{\partial x_{12}^{2}},
  \end{equation} 
   where $ \mu_{12}=\frac{m_{1} m_{2}}{(m_{1}+ m_{2})}$ is the reduced mass of particle $1$ and $2$. The Hamiltonian then becomes
\begin{eqnarray}
 {\hat H} = -\frac{1}{2 \mu_{12}}  \frac{\partial^{2}}{\partial x_{12}^{2}} + \frac{Z_{1} Z_{2}}{|x_{12}|}-\frac{1}{2 (m_{1}+m_{2})}  \frac{\partial^{2}}{\partial X_{12}^{2}}+ \sum_{i=3}^{N} \{
 Z_{i} (\frac{Z_{1}}{|x_{1i}|}+ \frac{Z_{2}}{|x_{2i}|})+v(x_{i})\} \nonumber \\
 +\sum_{i=3}^{N} \frac{1}{2 m_{i}}  \frac{\partial^{2}}{\partial x_{i}^{2}}+ \sum_{j>i=3}^{N} \frac{Z_{i} Z_{j}}{|x_{ij}|}+ v(x_{1})+ v(x_{2}).
 \end{eqnarray} 
 When particles $1$ and $2$ are within a small distance of each other( $0<|x_{12}|<\epsilon$), and all other particles are well separated, then there is only one
singularity in the Hamiltonian. Retaining  only the lower order terms in  $|x_{12}|$,  Eq. (8) reduces to
 \begin{equation}
[ -\frac{1}{2 \mu_{12}} \frac{\partial^{2}}{\partial x_{12}^{2}} + \frac{Z_{1} Z_{2}}{|x_{12}|}+ O(\epsilon ^{0})] \Psi (x_{1},x_{2},...,x_{N})=0,
\end{equation} 
 where $O(\epsilon ^{0})$ implies terms of order zero (constant),  one ( $|x_{12}|$ and $x_{12}$), and higher order in  $|x_{12}|$ and  $x_{12}$. 
 Note Eq.(9) is not an eigenvalue equation. In the limit $x_{1}\rightarrow x_{2}$ we write
 the wave function as
 \begin{equation}
  \Psi (x_{1},x_{2},...x_{N})= \Psi (x_{2},x_{2},x_{3}...,x_{N})+ \delta \Psi (x_{1},x_{2},...x_{N}),
\end{equation}
  where the term $\delta \Psi (x_{1},x_{2},...x_{N})$ vanishes at the singularity 
  $x_{1}=x_{2}$. From the differential equation Eq.(9) it follows that we need consider
  only terms of first order in $|x_{12}|$ and $x_{12}$. Thus, for $x_{1}\rightarrow x_{2}$
  we write the wavefunction as 
 \begin{equation}
  \Psi (x_{1},x_{2},...x_{N})= \Psi (x_{2},x_{2},x_{3}...,x_{N})+ |x_{12}|  B( x_{2},x _{3},...,x_{N}) + x_{12} \; C(x_{2},x_{3},...,x_{N})+ O(\epsilon ^{2}).
\end{equation}
 Substituting Eq. (11) into (9) we have 
 \begin{equation}
  -\frac{1}{2 \mu_{12}}  \; \frac{\partial^{2}|x_{12}|}{\partial x_{12}^{2}} B( x_{2},x _{3},...,x_{N}) + \frac{Z_{1} Z_{2}} {|x_{12}|} \;[\Psi (x_{2}, x_{2}, x_{3},...,x_{N})+Z_{1} Z_{2} B( x_{2},x _{3},...,x_{N})] =O(\epsilon^{0}).
 \end{equation}
 Since $\frac{\partial|x|}{\partial x}=sgn(x)$,  $\frac{\partial^{2}|x|}{\partial x^{2}}=2 \delta(x)$ so that Eq.(12) is
 \begin{equation}
 -\frac{1}{\mu_{12}} \delta(x_{12}) B(x_{2},x _{3},...,x_{N})+\frac{Z_{1} Z_{2}} {|x_{12}|} \;[\Psi (x_{2}, x_{2}, x_{3},...,x_{N})=O(\epsilon^{0}).
 \end{equation}
 In order for the singularities in each term of Eq.(13) to be cancelled, we must have
 $B(x_{2},x _{3},...,x_{N})=0$ and $\Psi (x_{2}, x_{2}, x_{3},...,x_{N})=0$. \textit{The latter proves that the particles cannot coalesce}. The node coalescence condition on the wavefunction  as 
 $x_{1}\rightarrow x_{2}$ is then
 \begin{equation}
   \Psi (x_{1}, x_{2}, x_{3},...,x_{N})= x_{12}  C(x_{2},x_{3},...,x_{N})+O(\epsilon ^{2}).
 \end{equation} 
 This condition is independent of the topology of the many-body system.
 The wavefunction of Eq.(14) is antisymmetric in the interchange of particles $1$ and $2$.
 For identical particles with antiparallel spin, it then follows that $C(x_{2},x_{3},...,x_{N})=0$. For identical particles with parallel spin, $C(x_{2},x_{3},...,x_{N})$ is not necessarily zero.\\

  We have  thus  proved rigorously that in $D=1$ dimension space the wave function of quantal particles
  interacting via the Coulomb force  in the presence of an arbitrary external force satisfies \textit{only} a \textit{node coalesence condition}. Hence these particles
  cannot coalesce and therefore cannot cross each other. They must consequently be ordered, and are as a result completely distinguishable . Thus,
  in one dimension the statistics of these interacting particles is of no significance.
  For non-interacting particles in one dimension, however, the energy spectrum differs
  depending on whether these particles are fermions or bosons. Consequently, as opposed to the $D=3$ high density  limit of the uniform electron gas \cite{10}, or of adiabatic coupling constant perturbation theory \cite{11}, the noninteracting system cannot be
  employed as the lowest-order in a perturbation theory of the interacting system. All
  the conclusions arrived at for the case of the Coulomb interaction are equally valid for the short-ranged screened-Coulomb(Yukawa) interaction.
  \\
  
  The case of the short-ranged delta-function interaction $\lambda_{12}\delta(x_{12})$ of interaction strength $\lambda_{12}$, which has been employed to obtain exactly solvable
  results\cite{12}, is different. Following the above steps, the corresponding coalescence
  condition for the wavefunction is derived as 
  \begin{equation}
  \Psi (x_{1},x_{2},...x_{N})= \Psi (x_{2},x_{2},x_{3}...,x_{N})(1+ \lambda_{12} \mu_{12} |x_{12}|)   + x_{12}  C(x_{2},x_{3},...,x_{N})+ O(\epsilon ^{2}).
\end{equation}
  Note that this condition is similar to the Coulomb interaction $D\geq 2$ dimension case described by Eq.(1). Thus, particles
  interacting via this hypothetical interaction can coalesce and  cross each other. As such,  the statistics of these particles are significant.\\

  In conclusion and for completeness, we note that in the literature of one-dimension
  systems\cite{1}, the ordering of particles is \textit{explicity} assumed. Thus, for example,
  quantal particles with harmonic external potential energy $v(x_{i}) = x_{i}^{2}$  and  centrifugal interaction potential energy $g/(x_{i} - x_{j} )^{2}$ are also \textit{assumed} \cite{13} not to coalesce, and therefore to be ordered and distinguishable. The wave function is then derived following this assumption. The reasons given for the assumption are the singular nature of the interaction at coalescence and the dimensionality of the problem.  However, as we have seen, the short-ranged delta-function interaction in one dimension is also singular but allows for coalescence.  Hence, the rationale for the assumption is not rigorous as it is for the Coulomb interaction case derived in the present work.  Furthermore, our results and conclusions are for \textit{arbitrary} external potential energy. The ability or lack thereof 
  of the particles to coalescence is not a function of whether the
  interaction is short-ranged or long-ranged. As noted above, particles
  interacting via the short-ranged screened-Coulomb interaction
  also cannot coalesce.  Additionally, we note that the Schrodinger equation for quantal particles in one-dimension having a \textit{combined} harmonic and centrifugal interaction potential energy ( $k( x_{1} - x_{2})^2/2 + g/(x_{1} - x_{2})^2 $)  but having \textit{no} external potential energy \cite{14} can be solved exactly. The wave function of these particles vanishes at coalescence, and thus they too are ordered and distinguishable. 
 We conclude by noting that node coalesence is fundamental  to the ordering of particles
 in one-dimension.
  \\

 This work was supported in part by the Research Foundation of the
 City University of New York.

\subsection{}
\subsubsection{}


\begin{references}
\bibitem{1} \textit{The Many-Body Problem, An Encyclopedia of Exactly Solved Models in One Dimension},  edited by D. C. Mattis, World Scientific 1993.
\bibitem{2}R. de L. Kroning and W. G. Penney, Proc. Roy. Soc.(London) \textbf{A 130},499(1931).
\bibitem{3}A. R. Goni, \textit{et al}, Phys. Rev. Lett. \textbf{67}, 3298 (1991); H. J. Schulz, Phys. Rev. Lett. \textbf{71},1864 (1993); B. Razaznejad, \textit{et al}, Phys. Rev. Lett \textbf{90}, 236803 (2003).
\bibitem{4}T. L. Einstein \textit{et al}, arXiv:cond-math/0012274v1 (14 Dec 2000).
\bibitem{5}X.-Y. Pan and V. Sahni, J. Chem. Phys.  (2003).
\bibitem{6}T. Kato, Commun. Pure Appl. Math. \textbf{10}, 151 (1957); W. A. Bingel, Z. Maturforsch. \textbf{18a}, 1249 (1963); R. T. Pack and W. B. Brown, J. Chem. Phys. \textbf{45},556 (1966); W. A. Bingel, Theoret. Chim. Acta. (Berl.) \textbf{8}, 54 (1967).
\bibitem{7} N. R. Kestner and O. Sinanoglu, Phys. Rev. \textbf{128}, 2687 (1962); S. Kais, D. R. Herschbach, and R. D. Levine, J. Chem. Phys. \textbf{91}, 7791, (1989); M. Taut, Phys. Rev. A \textbf{48}, 3561 (1993).
\bibitem{8}Z. Qian and V. Sahni, Phys. Rev. A \textbf{57}, 2527 (1998).
\bibitem{9} R. B. Laughlin, Rev. Mod. Phys. \textbf{71}, 863 (1998).
\bibitem{10} M. Gell-Mann and K. A. Brueckner, Phys. Rev. \textbf{106}, 364(1957).
\bibitem{11} A. G{\" o}rling and M. Levy, Phys. Rev. B \textbf{47}, 13105(1993).
\bibitem{12} A. A. Frost, J. Chem. Phys. \textbf{25}, 1150 (1956); J. McGuire, J. Math. Phys. \textbf{6}, 432 (1965); C. N. Yang, Phys. Rev. Lett.\textbf{ 71}, 1312 (1967).

\bibitem{13}B. Sutherland, J. Math. Phys. \textbf{12}, 246 (1971); ibid \textbf{12}, 251 (1971); Phys. Rev. A \textbf{4}, 2019 (1971).
\bibitem{14}F. Calogero, J. Math. Phys. \textbf{10}, 2191 (1969); ibid \textbf{12}, 419 (1971).

\end{references}
\end{document}